\documentclass{article}

\usepackage{arxiv}

\usepackage[utf8]{inputenc} 
\usepackage[T1]{fontenc}    
\usepackage{hyperref}       
\usepackage{url}            
\usepackage{booktabs}       
\usepackage{amsfonts}       
\usepackage{nicefrac}       
\usepackage{microtype}      
\usepackage{lipsum}		
\usepackage{graphicx}
\usepackage{natbib}
\usepackage{doi}
\usepackage{siunitx}
\usepackage{color}           
\usepackage{breakurl}        
\usepackage{multirow}
\usepackage{array}
\usepackage{wrapfig}
\usepackage{multirow}
\usepackage{tabularx}
\usepackage[pagewise]{lineno}
\DeclareUnicodeCharacter{2212}{-}
\DeclareUnicodeCharacter{2264}{-}

\chardef\us=`\_

\title{CNN-Based Deep Learning Model for Solar Wind Forecasting}

\author{ \href{https://orcid.org/0000-0002-5829-269}{\includegraphics[scale=0.06]{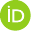}\hspace{1mm}Hemapriya Raju}\\
	Department of Astronomy, Astrophysics and Space Engineering\\
	Indian Institute of Technology Indore\\
    Indore 453552, India.\\
	\texttt{phd1901121008@iiti.ac.in} \\
	\And
	\href{https://orcid.org/0000-0003-4373-1631}{\includegraphics[scale=0.06]{orcid.pdf}\hspace{1mm}Saurabh Das} \\
	Department of Astronomy, Astrophysics and Space Engineering\\
	Indian Institute of Technology Indore\\
    Indore 453552, India.\\
	\texttt{saurabh.das@iiti.ac.in} \\
}

\renewcommand{\shorttitle}{CNN-Based Deep Learning Model for Solar Wind Forecasting}

\hypersetup{
pdftitle={A template for the arxiv style},
pdfsubject={q-bio.NC, q-bio.QM},
pdfauthor={Hemapriya Raju, Saurabh Das},
pdfkeywords={First keyword, Second keyword, More},
}

\begin{document}
\maketitle





\begin{abstract}
This article implements a Convolutional Neural Network (CNN)-based deep learning model for solar-wind prediction. Images from the \textit{Atmospheric Imaging Assembly} (AIA) at 193\,$\mathrm{Å}$ wavelength are used for training. Solar-wind speed is taken from the \textit{Advanced Composition Explorer} (ACE) located at the Lagrangian L\textsubscript{1} point. The proposed CNN architecture is designed from scratch for training with four years' data. The solar-wind has been ballistically traced back to the Sun assuming a constant speed during propagation, to obtain the corresponding coronal intensity data from AIA images. This forecasting scheme can predict the solar-wind speed well with a RMSE of 76.3\,$\pm$\,1.87\,\si{\km\per\s} and an overall correlation coefficient of 0.57\,$\pm$\,0.02 for the year 2018, while significantly outperforming benchmark models. The threat score for the model is around 0.46 in identifying the HSEs with zero false alarms.
\end{abstract}

%


\section{Introduction}
     \label{S-Introduction} 
The supersonic outflow of the plasma from the Sun's atmosphere, also known as solar-wind, has protons and electrons as its major constituents. Solar-wind travels through interplanetary medium and interacts with the Earth's magnetic field causing major disturbances and variations in the space environment. These changes in the space environment between Sun and Earth are referred to as space-weather \citep{Schwenn2006}. These disturbances show a 27-day re-occurrence period, which can be easily correlated with the solar rotation \citep{Bartels1932,SargentIII1985}.\\
Geomagnetic storms are mainly caused by fast wind ranging from 400\,--\,800 \,\si{\km\per\s} and can reach up to 1500\,\si{\km\per\s} \citep{Sheeley1976}. Solar storms can also lead to increased drag on satellites. For instance, \textit{Skylab} reentered the Earth's atmosphere earlier than expected due to atmospheric drag caused by increased solar activity (\url{sdo.gsfc.nasa.gov/mission/spaceweather.php}). Geomagnetic storms may also cause electric-power disruption and blackouts on a large scale. They also disturb high-frequency communication and navigation systems, as the density of the ionosphere varies abruptly, causing the satellite signals to pass through unexpected propagation paths \citep{angeo-27-2101-2009}. Thus understanding the origin of solar-wind and its accurate prediction helps in evaluating the nature of geomagnetic storms and provides the necessary time for preparedness.\\
Most of these adverse geomagnetic storms occur due to coronal mass ejections (CMEs) \citep{doi:10.1029/93JA02867,doi:10.1029/2007JA012744,Tsurutani2020}. CMEs commence suddenly and can be more often seen during the solar maximum during peak of solar activity \citep{Schwenn2006}. But, some of the space-weather disturbances are long-lasting, stable, and recurrent. These types of storms can be well noticed during the descending phase of the solar cycle, where solar activity will be at its minimal. These disturbances are mostly associated with coronal holes (CHs) \citep{HANSEN1976381}.\\
CHs appear visibly dark when observed in extreme ultraviolet (EUV) and UV radiation due to its reduced temperature. The principal source of fast wind during solar minima is the steady and long-lasting CHs. A fast wind can be characterized with velocity greater than 400\,\si{\km\per\s}, whereas a slow wind has a velocity less than 400\,\si{\km\per\s} \citep{Gosling1999}. The commencement of high speed streams (HSS) \citep{Krieger1973} can be seen as abrupt changes in wind speed from slow to fast. As CHs are primarily responsible for the persistent, periodic, and gradual onset of HSS, it will be useful to predict this recurrent solar-wind speed \citep{Neupert1974}.\\
\cite{Krieger1973} initially found that there is a correlation between recurring HSS and the position of the coronal hole in the descending phase of a solar cycle. This was backed by the similarity between the magnetic polarity found in the CH and the polarity of the solar-wind near Earth. Later, \cite{Harvey1979} proposed that the stable CH patterns and its magnetic field can be associated with reappearing HSS measured near Earth at regular intervals, thus signifying the nature of CH in the declining phase of the solar cycle.\\
Based on these observations and theoretical understanding of the phenomena, currently, there are several empirical and physical models laid out for the prediction of HSS. The Wang‐-Sheeley--Arge (WSA) model is the leading physical and empirical model that helps in predicting the interplanetary magnetic field (IMF) polarity and background solar-wind speed near Earth using 22 years of solar observations. This model uses the negative correlation between the flux-tube expansion factor in CHs and the solar-wind speed. The WSA model is used by a number of agencies including NOAA’s Space Weather Prediction Center (SWPC) for forecasting near-Earth solar-wind conditions \citep{1990ApJ...355..726W}.\\
Building on the presence of a relation between CH area and solar-wind speed, \cite{Rotter2012} found a high correlation of 0.78, while CH area had a vague relationship with other variables such as temperature [$T$] and magnetic field [$B$].
\cite{Owens2013} presented a 27-day persistence model (PS27). Repeated solar-wind source structures for each solar rotation are studied for forecasting solar-wind speed based on 27-day periodicity.
\cite{Rotter2015} further improved the empirical model based on the CH areas extracted from the \textit{Solar Dynamics Observatory}  (SDO)/\textit{Atmospheric Imaging Assembly} (AIA) images \citep{Lemen2012,2012SoPh..275....3P}. Utilizing the relation of fractional CH area obtained from the meridional slices of AIA images with solar-wind speed, the model forecasts up to four days prior to the occurrence. For this purpose, the dataset of 2011\,--\,2013 is prepared with cadence of one hour and they were able to obtain correlation of 0.60, with 80\,\% of the peaks predicted.
\cite{Tokumaru2017} presented an empirical relationship between the square root of CH area [$A^{1/2} $] and solar-wind speed. High correlation (0.70) has been observed between those two variables investigated between the period of 1995 and 2011.
Recently, \cite{Bu2019} presented an empirical model for solar-wind speed prediction by utilizing pixel values in the coronal-hole region of the AIA image. Introducing a new input parameter P\textsubscript{CH}, the model improved the forecast accuracy when compared to 27-day persistence model for the period of 2011\,--\,2018, with temporal cadence of one hour. The method identified 73\,\% of the peaks correctly with the threat score at 0.61, which is used to measure the model's performance and root mean squared error (RMSE) of 102.9\,\si{\km\per\s}.\\
Apart from empirical and physical models, machine learning (ML) as well as deep learning (DL) models are emerging as increasingly popular domains for research in solar physics. \cite{Liu2011} applied the support vector regression (SVR) algorithm to predict the solar-wind velocity with input as four 27-day period of solar-wind data with a cadence of one hour. The model had been trained with nine years of solar-wind data from 1998 to 2006. Although this model has prediction accuracy of 97.9\,\%, it was able to predict only one to three hours ahead of the occurrence of the solar-wind.
\cite{Yang2018} employed artificial neural network (ANN) for forecasting solar-wind speed. The model is a hybrid framework using several types of theoretical and observational information as input. Different combinations of inputs are tested including individual variables, to select an ideal model for forecast. The results are analysed for the period from 2007 to 2016 with a cadence of one hour, and it was able to predict 68.2\,\% of the observed high speed events (HSEs) with a correlation of 0.74, RMSE of 68\,\si{\km\per\s}, and threat score of 0.55.
\cite{Bailey2021} implemented another ML model based on gradient boosting regressors (GBR) in improving ambient solar-wind predictions. In addition to the regular variables for training, to account for uncertainties in the magnetic field, this model utilizes all 12 air force data assimilative photospheric flux transport (ADAPT) realizations, which provides a possible magnetic field on the far side of the Sun. Their model has outperformed existing models and provides forecasts with a lead time of 4.5 days.\\
Currently, deep learning is revolutionizing the world from all corners of the industry (e.g. healthcare, social media etc.) with its exemplary performance. Deep learning refers to a network of algorithms used to learn sophisticated non-linear relationship between data, as opposed to the single-algorithm approach used in machine learning. Prior to deep learning, non-neural-network based models were able to achieve only 25.7\,\% in top five classification error rate on the ILSVRC 2012 model dataset. AlexNet \citep{Krizhevsky2012} was the first DL model to achieve 15.3\,\% in top five classification error rate on the same data set, thus showing the efficacy of deep-neural-network-based systems \citep{DBLP:journals/corr/LiuRPCKLKWC16}. 
Most of the state-of-art deep learning architectures use hierarchical feature learning known as the convolutional neural network (CNN) for computer vision and visual pattern recognition. Present-day CNN frameworks are structured to be deep and large with many hidden layers, making them ﬂexible in learning a wide class of examples simultaneously from the available data.\\
The recent implementation of deep learning in solar physics has shown very encouraging results. Very recently \cite{Upendran2020} had proposed a DL model called Windnet for predicting solar-wind speed. The architecture has a combination of pretrained GoogleNet with Long Short-Term Memory (LSTM) for regressing against daily averaged solar-wind speed. Partitioning of datasets into batches, with each batch consisting 20 days of contiguous data, is used for training and testing. Two hyperparameters are used, namely history [$H$], which represents number of days of input image needed for one prediction, and delay [$D$] representing the time lag by which solar-wind speed is mapped to the AIA images during training. Based on the combination of these two parameters, an optimum model is selected for prediction. The model was trained and tested for the period 2011\,--\,2018 and had obtained an overall correlation of 0.55$\,\pm\,0.03 $ with maximum threat score of 0.35. \\
In this article, we have tried to develop a deep-learning model for solar-wind speed predictions using CNN, similar to the \cite{Upendran2020} model. Four years of data from SDO/AIA images are used for developing the model. Similar to the \cite{Upendran2020} model, all pixel values are used instead of derived CH parameters. This preserves the information content that is other wise lost in identifying features. However, we eliminated the need for time-dependent LSTM as employed by \cite{Upendran2020}, by introducing an adaptive time-delay method. Moreover, \cite{Upendran2020} used cross validation for evaluation of their model for the period 2011\,--\,2018, but it is better to have a test set spaced far apart from the training set to prevent the influence of possible recurrence of structures that might have 27-day persistence. \cite{Yang2018} on the other hand, used a complete year as a test set. We followed a similar data separation and used the entire year of 2018 as the test set. Testing on a separate dataset of an entirely different year is essential to evaluate the generalization capability of the model.\\
The key issues that are addressed in the present work are as follows: 
 \begin{itemize}
    \item During the training phase, a heuristic approach is introduced to enhance the performance of CNN-based deep-learning model by ballistic back tracing of solar-wind speed to obtain the coronal image closest to the source region of the solar-wind.\\
    \item Proposed CNN model replaces the time-dependent architecture by introducing an adaptive delay method. This makes our architecture simple, less complex, and computationally light weight.\\
    \item To evaluate the generalization capability of the proposed model, the trained model is evaluated on a completely independent year-long dataset.\\
\end{itemize}
The organization of the article is as follows: Section 2 gives a comprehensive overview of CNN with the proposed architecture and hyperparameters. Section 3 describes the data source, preprocessing, training of the CNN model, and the proposed method of adaptive time-delay. In Section 4, discussion on training and validation loss curves, model uncertainty, validation of the model with a few benchmark models on continuous parameters, event-based approach, and visualization of CNN working are given. Section 5 presents discussion on challenges faced and scope for improvement in the model's performance. Section 6 presents a conclusion of the present work and improvements that can be made further beyond this article.
\section{Method}
CNN is the most successful deep-learning algorithm for extracting features of the image at good resolution by assigning weights. These features are turned into complex features at a coarser resolution, as the network goes deeper. The architecture of CNN can be divided into three layers. The initial layer is convolutional. It extracts features by carrying out convolving images with weights known as a kernel, which are initialized randomly. The kernel slides over the image with a certain stride value, thereby extracting the low-level features such as shapes and edges in the initial layer. This final output obtained at each layer, after applying these kernels, is known as feature maps. To extract distinct features with different weights, we can vary the number of kernels according to the model's requirement. Having more convolutional layers helps in extracting high-level features.\\
Each convolutional layer along with the activation function is immediately followed by a pooling layer. In this study, the activation layer used in the hidden layers is Rectified Linear Units (ReLU) having its range from zero to infinity as shown in Equation 1:
\begin{equation}
    \textit{R(z)}=\text{max(0,\textit{z})}
\end{equation}
Here, the sub-sampling of the image is done by operation of max pooling to decrease the spatial size of the resulting convolved image, and thus the computational power required for processing the data is minimized. Two types of pooling are used in practice, max and average pooling, which return maximum value and average value for the respective portion of the image slided over by the kernel. Using these two layers in a repeated manner deepens the network and helps in extracting significant features representing the image. Once the features are extracted, the image is flattened to a column vector and fed to a third layer called the fully connected layer, where all inputs and outputs are connected across each other in all layers.\\
From the feature map generated, the appropriate output is activated by an activation function, which depends on the kind of problem in hand, such as regression or classification. Since the current goal is to do a regression of solar-wind speed from solar EUV images, we have used a linear activation function.
\subsection{Proposed CNN Architecture}
In this work, a custom CNN model is designed to predict solar-wind speed from AIA images. The schematic diagram of the designed CNN architecture is shown in Figure 1.
\begin{figure}[h!]
  \centering
  \centerline{\includegraphics[width=\columnwidth,clip=]{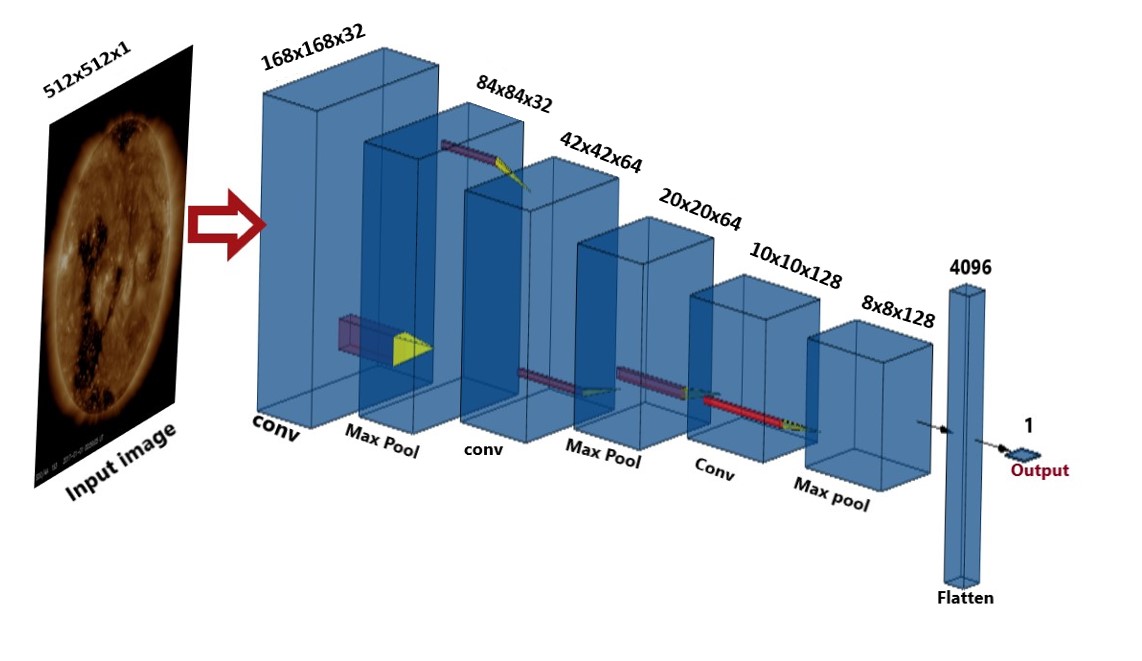}
              }
  \caption{Proposed CNN Architecture.}
  \label{fig:Figure 1}
\end{figure}
The proposed CNN model has three convolutional layers, with maxpooling layers at each end of the convolutional layers, followed by two fully connected layers.\\
The layers and parameters used in the designed CNN architecture are shown in Table 1. The activation function of the output layer is taken as linear as it is a regression problem. Since we have a limited number of images, we created our network not too deep to avoid overfitting. The network is set to train initially for 20 epochs, with callbacks for early stopping to monitor validation error.\\
\begin{table}[h!]
\caption{CNN Layers and Kernel size parameters. 
}
\label{tab5}
\begin{tabular}{lc}     
  \hline                   
CNN Layers & Kernel size \\
  \hline
 Conv1 & 9$\times$9 \\
 Pooling & 2$\times$2 \\
 Conv2 & 2$\times$2 \\
 Pooling & 2$\times$2 \\
 Conv3 & 2$\times$2 \\
 Pooling & 3$\times$3 \\
 Fully & 4096\\
 Fully & 1\\
  \hline
\end{tabular}
\end{table}
We also have Batch Normalization which, in short, normalizes the output from a previous activation layer by subtracting the mean of that batch and dividing by standard deviation. Batch Normalization helps in increasing the stability of a neural network \citep{NEURIPS2018_36072923}, as standardizing inputs for every minibatch  reduces the internal covariance shift. It is added at the end of the convolutional layer in this proposed architecture. A neural network is comprised of nodes, where one or more inputs along with weights are passed through an activation function, by which the output to the next layer is determined. In dropouts \citep{JMLR:v15:srivastava14a}, a few nodes are randomly dropped out or ignored each time while the network is trained. The use of dropouts acts as a regularization technique and helps in reducing overfitting. Dropouts can also be used as a Bayesian approximation to determine the model uncertainty \citep{pmlr-v48-gal16}. Dropout is added in the dense layer and the model is compiled with the Adam optimizer. The Adam optimizer \citep{DBLP:journals/corr/KingmaB14} works with an adaptive learning rate for every training step and updates different parameters with individual learning rates. During the model compilation, we set the maximum learning rate as \(10^{-4}\).
\subsection{Hyperparameters}
For a neural network, there are hyperparameters, which need to be identified for a better performance of the model.
Batch size is the number of training inputs utilized in one iteration and it can determine the model’s performance. We observed that a batch size of 128 gave good predictions while batch size of 32 and 64 resulted in poor predictions.
\begin{table}[h!]
    \caption{Hyperparameters tuned in the proposed CNN model} 
\label{tab1}
\begin{tabular}{lc}     
  \hline                   
Hyperparameter & Values \\
  \hline
Normalization& Batch \\
Dropouts& 0.3 \\
Number of hidden layers& 3 \\
Number of filters in Conv layer 1 &32\\
Number of filters in Conv layer 2 &64\\
Number of filters in Conv layer 3 &128\\
Batch size& 128 \\
epochs & 7\\
learning rate & 0.0001\\
  \hline
\end{tabular}
\end{table}
Batch Normalization helped in reducing overfitting in our model as it has a regularization effect. Fixed dropout of 0.3 also acts as a hyperparameter, which was decided using the trial and error method. \\
We decided to keep the number of convolutional layers as three, because two-layered convolutional models were too shallow to learn, resulting in the underfitting of the data. As there is a smaller number of images (around 16k) in training and validation, we observed that a too deep layered convolutional model also results in reduced performance compared to the three-layered convolutional model, as there is a trade-off between the number of hidden layers and the size of training data. A quick overview of the hyperparameters is given in Table 2 for reference.\\
We use mean squared error (MSE) as the error function in this article, given by Equation 2:\\
\begin{equation}
    \text{MSE} = \mathnormal{\frac{1}{n}\sum_{i=1}^{n}(y_{i} - x_{i})^{2}}
\end{equation}
The training of our network took 15 minutes on a server having an AMD EPYC 7401 24-Core Processor with 256\,GB RAM and 96 AMD processors, while it took less than two minutes to train with a graphics processing unit (GPU): GeForce RTX 2060 6\,GB RAM. The GPU has 1920 NVIDIA CUDA cores for parallel processing, with more cores equating to better performance. This GPU comes with Turing architecture, which has dedicated tensor cores, ideal for AI computing.
\section{Data Source and Pre-Processing}
\subsection{SDO/AIA Images}
\subsubsection{Data Source of AIA Images}
The SDO was launched to understand the dynamics of solar activity and its impacts on the Earth \citep{2012SoPh..275....3P}. SDO is intended to assist us with understanding the Sun's effect on near-Earth space by exploring the solar atmosphere in different wavelengths at the same time. SDO consists of three scientific instruments among which AIA \citep{Lemen2012} has been providing continuous full-disk observations of Sun's chromosphere and corona in seven EUV channels. As per previous studies \citep{Yang2011}, 193\,\si{\angstrom} is more sensitive to coronal holes. The curated ML dataset \citep{Galvez_2019} containing AIA images of that particular band has been collected from the Stanford digital repository (\url{https://purl.stanford.edu/km388vz4371}). Our training set consists of images taken from AIA of four years and a testing dataset of one year with cadence of six minutes and 512$\times$512 spatial coverage.
\subsubsection{Data Gaps and Mitigation}
Missing images are replaced by the images from the original dataset that fall in the same hour. If no such images are found in that hour, then missing images were replaced by the images present in the nearest hour. In total we had 46 missing images for 2013, 283 missing images for 2014, 66 missing images for 2015, 76 missing images for 2017, and 95 missing images for 2018. A large portion of images were unavailable for 2011, 2012, and 2016. Hence we chose to ignore those years to prevent our dataset from having more duplicate images.
\subsubsection{Preprocessing of Images}
The pixel values had a high dynamic range of values, hence we performed log scaling to convert the pixel values into an eight-bit integer in the range of 0\,--\,255. We tried with various ranges to determine maximum and minimum threshold. For this study, log scaling with an upper limit of 5000 and lower limit of 125 gave good results, similar to the work of \cite{Upendran2020}. Studies can be done further to explore this dynamic range. Finally, we sampled the image at two-hour cadence, yielding around 16\,k images for training and validation.\\
\subsection{Solar-Wind Data}
\subsubsection{Data Source for Solar Wind}
These images are trained against solar-wind speed as the target variable, taken from the \textit{Advance Composition Explorer} (ACE) repository. NASA's ACE was designed to analyze particles of solar and interplanetary origins. The spacecraft is located at Sun\,--\,Earth Lagrange L\textsubscript{1} point and gives space-weather reports and warnings of geomagnetic storms one hour ahead of arrival at Earth (\url{www.swpc.noaa.gov/products/ace-real-time-solar-wind}). The solar-wind speed is taken with two-hour cadence from the ACE repository (\url{www.srl.caltech.edu/ACE/ASC/level2.html}.)
\subsubsection{Data Gaps}
The data from 2013 has 50 missing values, 10 missing values were present in 2014, 2 missing values were present in 2015, and 18 values were absent in 2017 data. The test dataset of 2018 had 51 missing values.
 \subsubsection{Preprocessing of Solar-Wind Data} 
 Preprocessing of solar-wind data includes checking for missing values and replacing them with adjacent values. The final temporal cadence of the data is maintained the same as the original cadence.\\
\subsection{Dataset Split}
Images and solar-wind speed for 2013\,--\,15 and 2017 are pre-processed and randomly split as training and validation dataset in the ratio of 80\,\% and 20\,\%. A separate dataset for 2018 is used for testing the performance of the developed CNN model. The histogram distribution of solar-wind speed for training, validation, and test dataset are shown in Figure 2. From the figure, we can see that the distribution of training, validation, and test dataset are similar.
\begin{figure}[h!]
  \centering
  \includegraphics[width=\columnwidth]{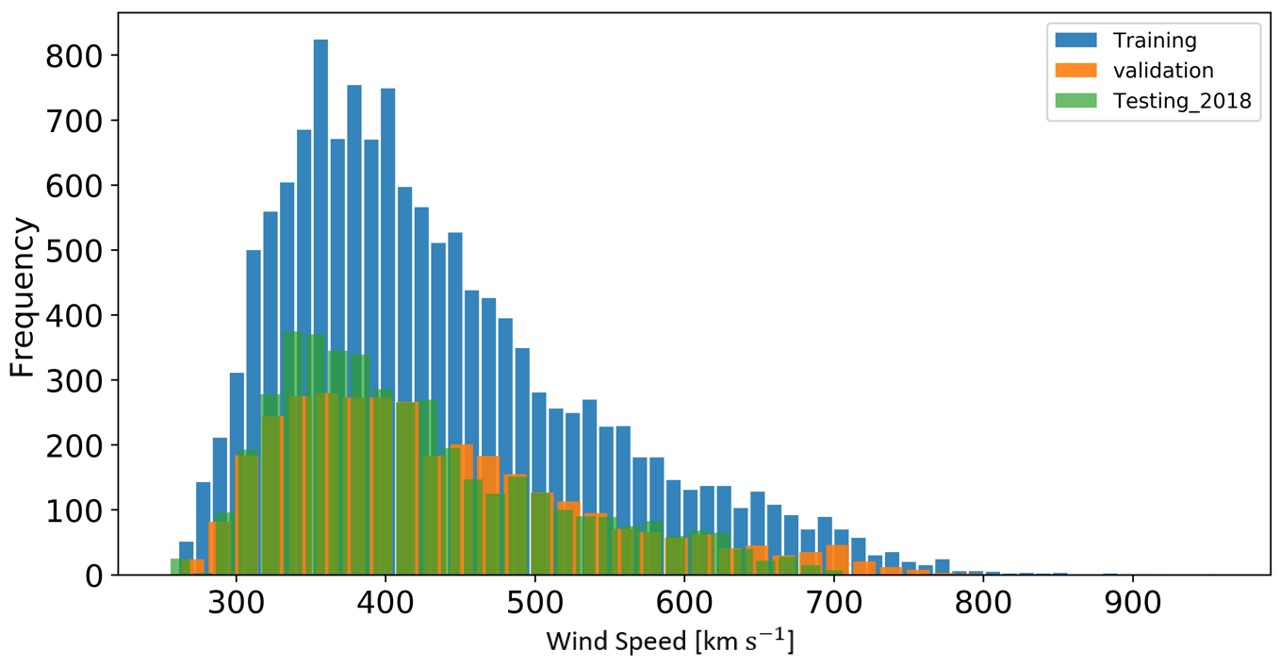}
  \caption{Distribution of solar-wind speed of training,validation, and test dataset.}
   \label{fig:Figure 3}
\end{figure}
\subsection{Time Delay Mapping}
The measured speed must be ballistically back traced to obtain the AIA image corresponding to the source region of the measured solar-wind. In general, fast wind usually takes around three to four days and slow wind takes greater than five days to reach from Sun to Earth. In previous works, \citep{1990ApJ...355..726W,Vrsnak2007} an average of a four-day delay is taken for prediction. Average time delay while mapping AIA images with solar-wind speed might lead to increased inaccuracy in the model. Thus, time delay at L\textsubscript{1} calculated as,
\begin{equation}
    \textit{Time delay at L\textsubscript{1}} = \textit{1\,AU}/\textit{observed wind speed}
\end{equation} 
and respective AIA images are mapped back with that particular time delay during training. A typical time series of AIA images and its corresponding time-lagged wind speed is shown in Figure 3. 
\begin{figure}[h!]
  \centering
  \includegraphics[width=\columnwidth]{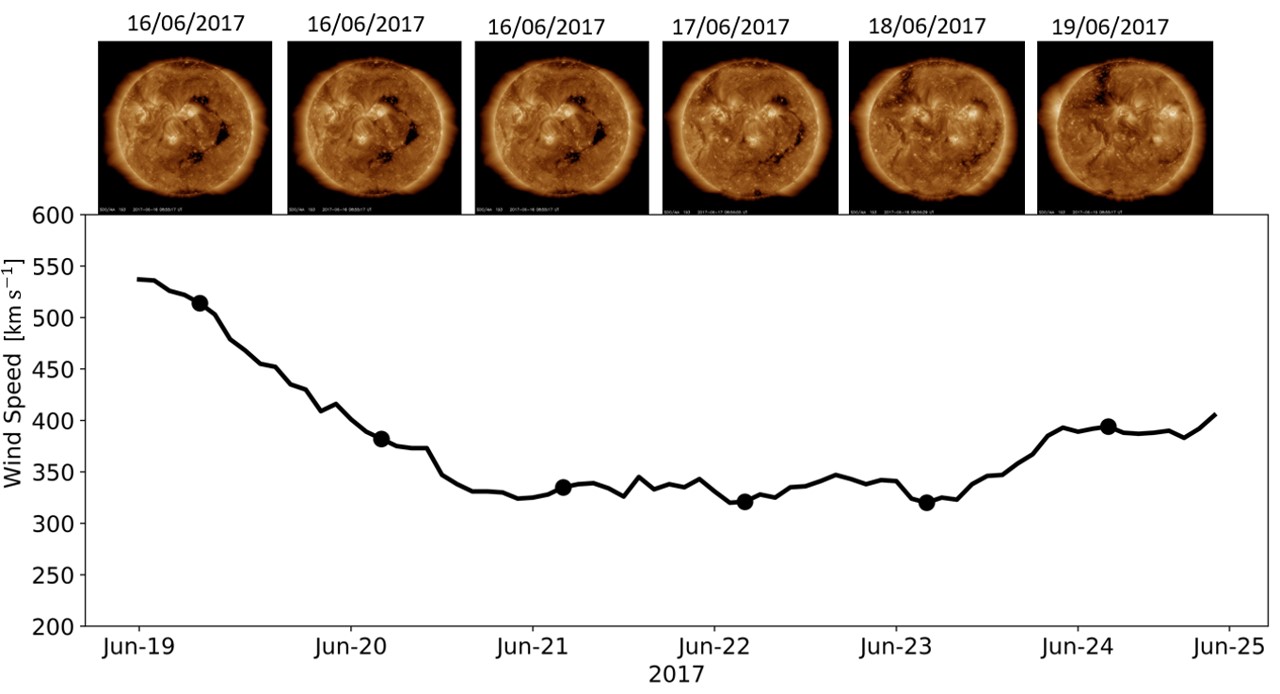}
  \caption{Time series of AIA Images and time lagged wind speed consisting of both fast and slow wind is marked for a few days in June 2017. }
  \label{fig:Figure 2}
\end{figure}
As a result, the training dataset may contain instances where multiple wind speeds were backtracked to the same AIA image as observed in Figure 3. The time series of AIA images is resampled as follows:\\
\begin{equation}
    \text{new series\textit{[i]}=old series[\textit{i}-(time delay)]}
\end{equation} 
We stress here that the adaptive time-delay method is purely a heuristic assumption for training the model. Hence it should not be used for understanding the propagation of the solar-wind, which in reality flows with varying velocity, and follows an archimedian spiral \citep{Poduval2014}. However it turns out that this simple assumption resulted in good performance of the model.\\
As already discussed, the model is developed based on the dataset of 2013\,--\,15 and 2017. The model is tested against entirely different dataset of the year 2018 to ensure that the model generalizes well for unseen data. The SDO/AIA images for the year 2018 are preprocessed in the same way as training datasets.
\section{Results}
The source code of the model has been made open source and is accessible online via github (github.com/Hemapriya-iit/CNN-based-solar-wind-prediction). In this section, the results of the proposed CNN model for the whole year 2018 are discussed and its statistics are represented with a temporal resolution of two hours.\\
\subsection{Model Validation by Continuous Variables}
To get a basic idea of the model's performance, metrics such as RMSE and Pearson correlation are evaluated.
RMSE can be calculated as the square root of the arithmetic mean taken for the squared difference between observed and predicted values, as shown in Equation 5. RMSE for the CNN model is around 76.30$\,\pm$\,1.87\,\si{\km\per\s}.
\begin{equation}
    \text{RMSE} = \mathnormal{\sqrt{\frac{1}{n}\sum_{i=1}^{n}(\text{y\textsubscript{predicted}} - \text{y\textsubscript{observed}})^{2}}}
\end{equation}
The Pearson correlation, calculated between -1 and +1, describes the relationship between two time series in this context. The Pearson correlation coefficient [$r$] is calculated as the covariance of the time series divided by the product of the standard deviation, as shown in Equation 6:\\
\begin{equation}
 r = \frac{n \sum{x_iy_i}-\sum{x_i}\sum{y_i}}{\sqrt{ n \sum{x_i^2}-(\sum{x_i})^2} \sqrt{ n\sum{y_i^2}-(\sum{y_i})^2 }}
 \end{equation}
The overall correlation of the CNN model ranges around 0.57\,$\pm$\,0.02. We have performed the averaging of data from 100 trials in Fisher’s Z-space and later transformed it to correlation. A positive correlation suggests that the two time series are positively linearly related.\\
A simple 2D histogram distribution of predicted wind speed by CNN compared with actual wind speed for train, validation, and test datasets with its respective correlation coefficient can be seen in Figure 4. It can be seen that the model performs better for the validation dataset also, which comprises of dataset from both solar maximum and descending phase.
\begin{figure}[h!]
\centering
  \centerline{\includegraphics[width=\columnwidth,clip=]{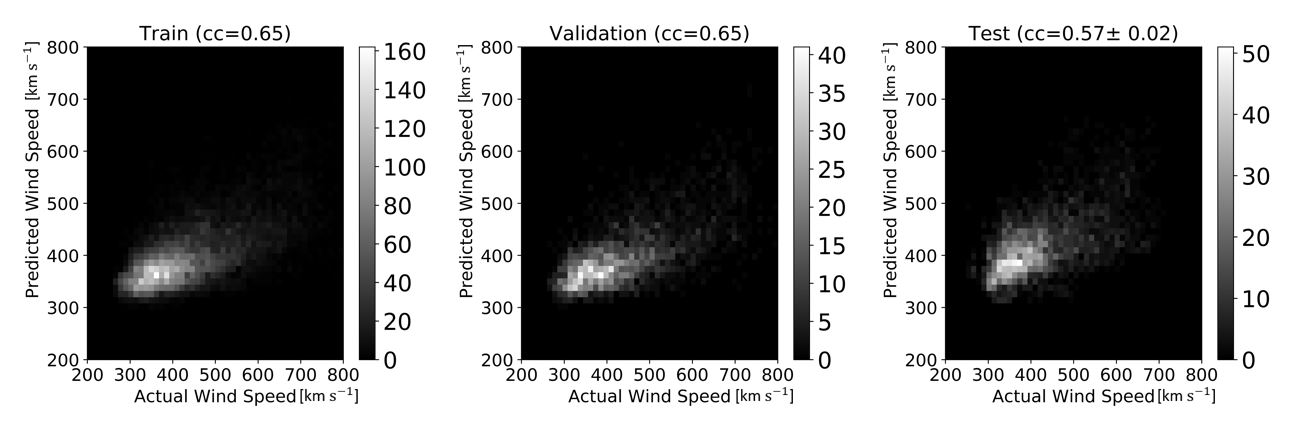}
              }
              \caption{2D histogram distribution of observed versus predicted wind speed individually for train, validation, and test dataset with its respective correlation mentioned at the top of each figure. The color bar indicates the frequency of occurrence with each bin of 12\,\si{\km\per\s}.}
  \label{fig:asc}
\end{figure}

\subsubsection{Loss Curves}
In Figure 5, the mean squared error loss curve during model training has been shown. It can be noted that the both validation loss and training loss converge fast around three epochs, beyond which they reduce slowly with occasional increase in the loss.
\begin{figure}[h!]
    \centering
    \includegraphics[width=\columnwidth]{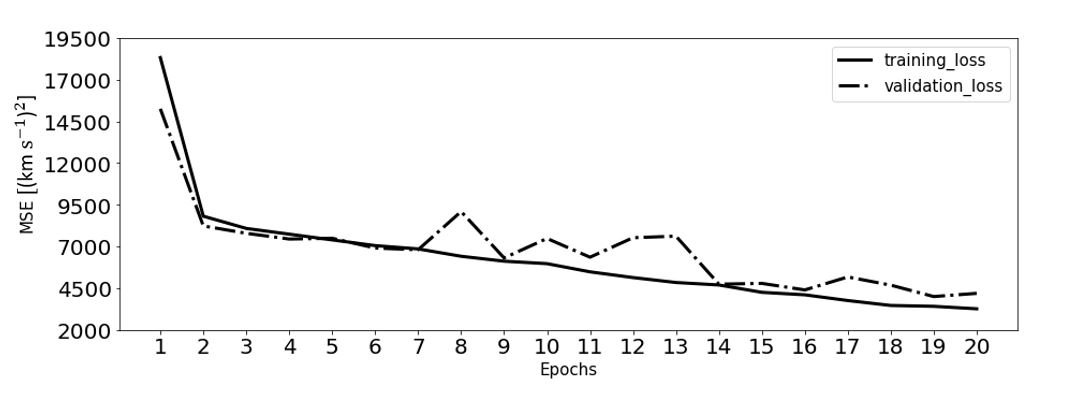}
    \caption{Variation of training loss (red) and validation loss (green) for our CNN model during training.}
    \label{fig:asc}
\end{figure}
In general, when we trained the network for various epochs, the performance of the model varies. The CNN model gives best performance when trained with five to seven epochs. We have used early stop as a callback parameter, where it monitors the validation loss. If validation loss increases consecutively beyond three epochs, then the training is set to stop by itself. This callback parameter helped us to stop our training around the seventh epoch. Further we did analysis from 3 to 20 epochs individually, without using callback parameter, which also shows that the network performance was better between the fifth and seventh epoch. Table 3 shows the variations in the statistical measures for different epochs for validation and test dataset. It can be seen that after the seventh epoch both validation and test loss increases.\\
\begin{table}[h!]
    \caption{Comparison of statisical measurements for different number of epochs}
\label{tab1}
\begin{tabular}{cp{2cm}p{1.5cm}}

  \hline
Epochs & Validation RMSE [\si{\km\per\s}] & Test RMSE [\si{\km\per\s}]  \\
  \hline
3 & 92 & 83 \\

5 & 89 & 76 \\
  
6 & 85 & 76 \\
  
7 & 82 & 76 \\
 
8 & 81 & 78 \\

9 & 94 & 80 \\

10 & 77 & 79 \\

11 & 85 & 79 \\
  \hline
\end{tabular}
\end{table}
\subsubsection{Uncertainty in Model Predictions}
The validation of a model must account for the uncertainties present in the predicted data during model processing. Quantifying the uncertainty is critical to understand the extent of the model’s actual success. Inherent uncertainty present in the observed data itself is known as aleatoric uncertainty. Epistemic uncertainty also known as systematic uncertainty, arises due to limited information or lack of knowledge in understanding a particular process.\\
Aleatoric uncertainty, here should be understood as the variation in the observed averaged solar-wind speed. 12-minute cadence data is used to calculate the average for two-hour data. The measurement error is calculated by observing the variations within those two hours. The measurement error  with 1$\sigma$ for a few days are shown in Figure 6. The distribution of the measurement error in the solar-wind speed for 2018 can be seen in Figure 7.\\
\begin{figure}[h!]
    \centering
    \includegraphics[width=\columnwidth]{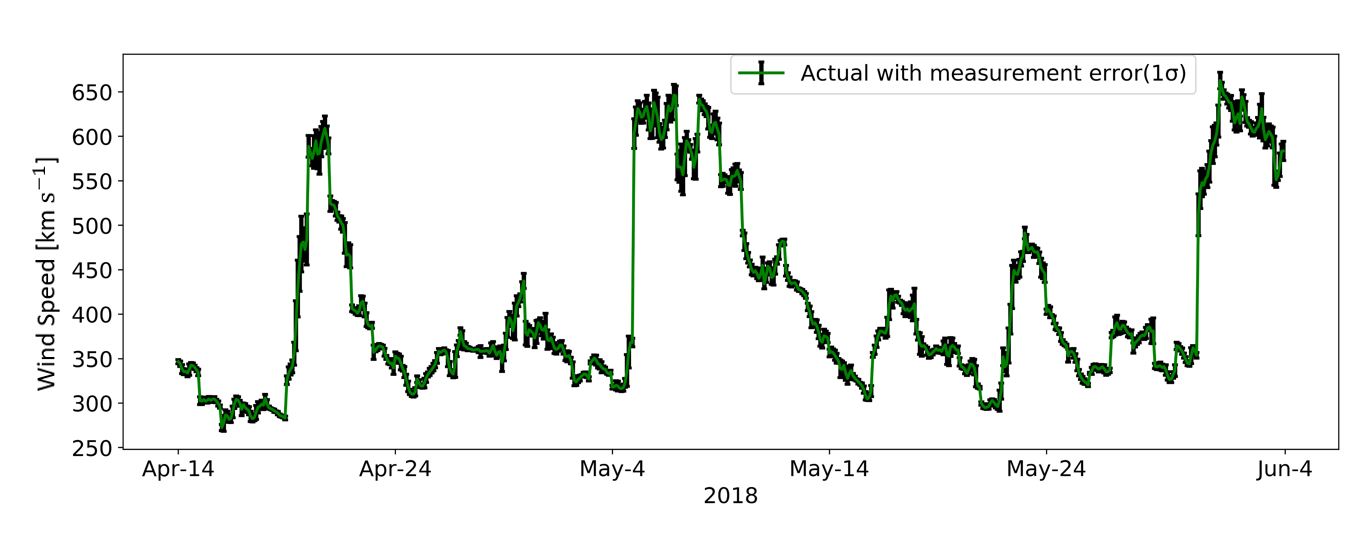}
    \caption{Time series of two-hour averaged actual wind speed with measurement error 1$\sigma$ for the month of April\,--\,May 2018.}
\end{figure}
\begin{figure}[h!]
    \centering
    \includegraphics[width=\columnwidth]{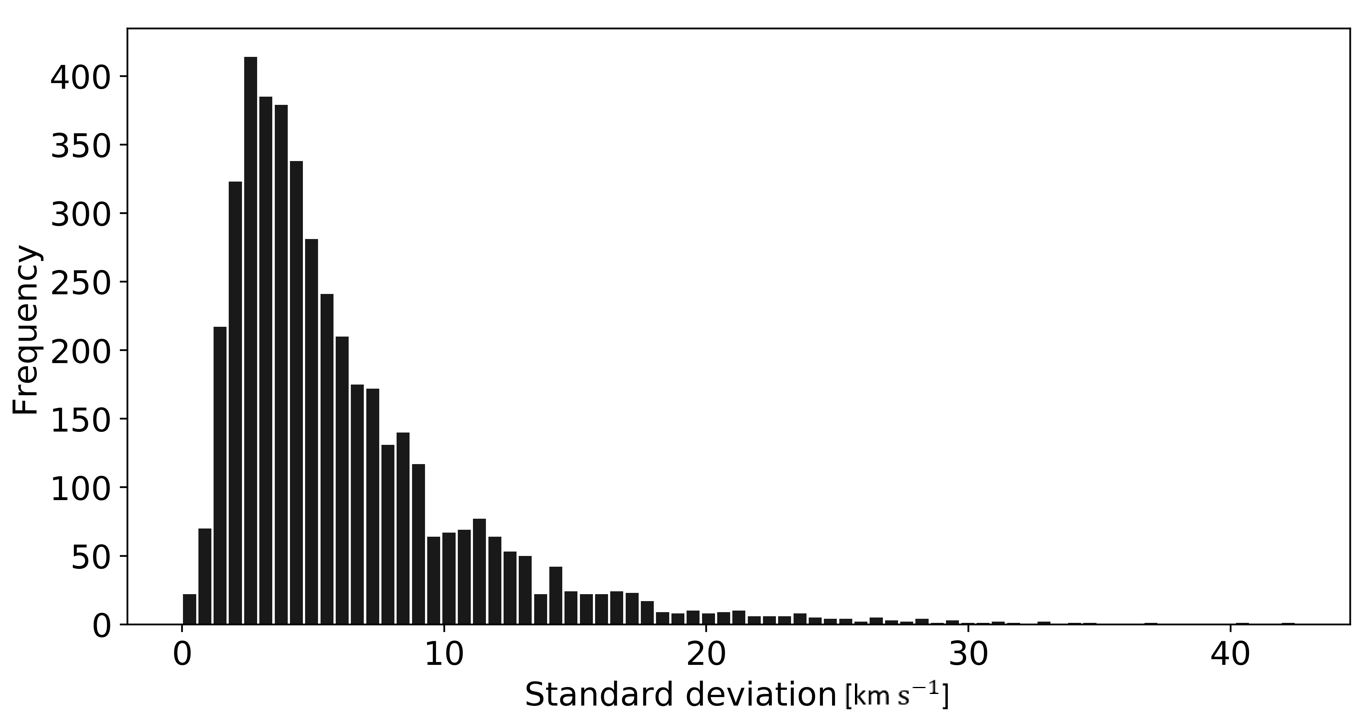}
    \caption{Histogram of the variation in measured solar-wind speed averaged at two-hour cadence, for whole year 2018.}
\end{figure}
Epistemic uncertainty in the current context can be understood as the extent of variation in the forecasted solar-wind speed as a result of unexplored areas in the weight space. This is quantified by performing Bayesian approximation using dropouts \citep{pmlr-v48-gal16}. Dropouts appear to be a simple way of converting the deep-learning models to Bayesian models, without any significant change in the model structure.
\begin{figure}[h!]
     \centering
         \includegraphics[width=\columnwidth,clip=]{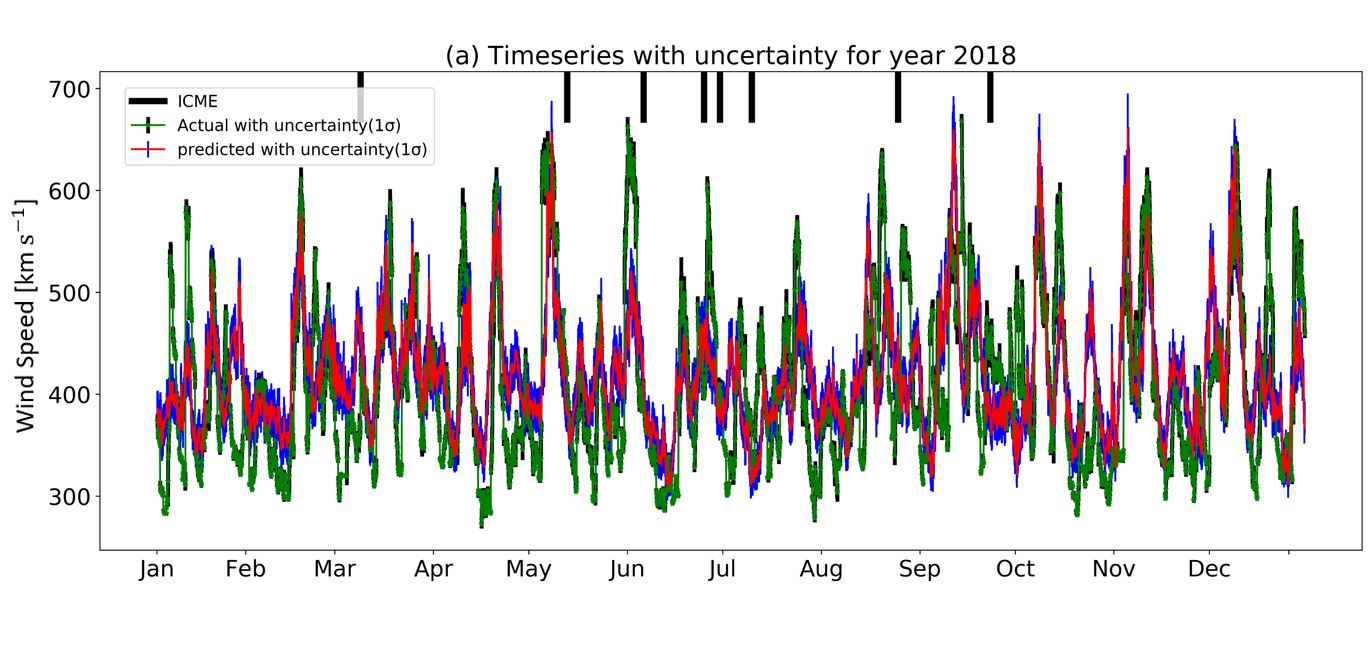}
         \includegraphics[width=\columnwidth,clip=]{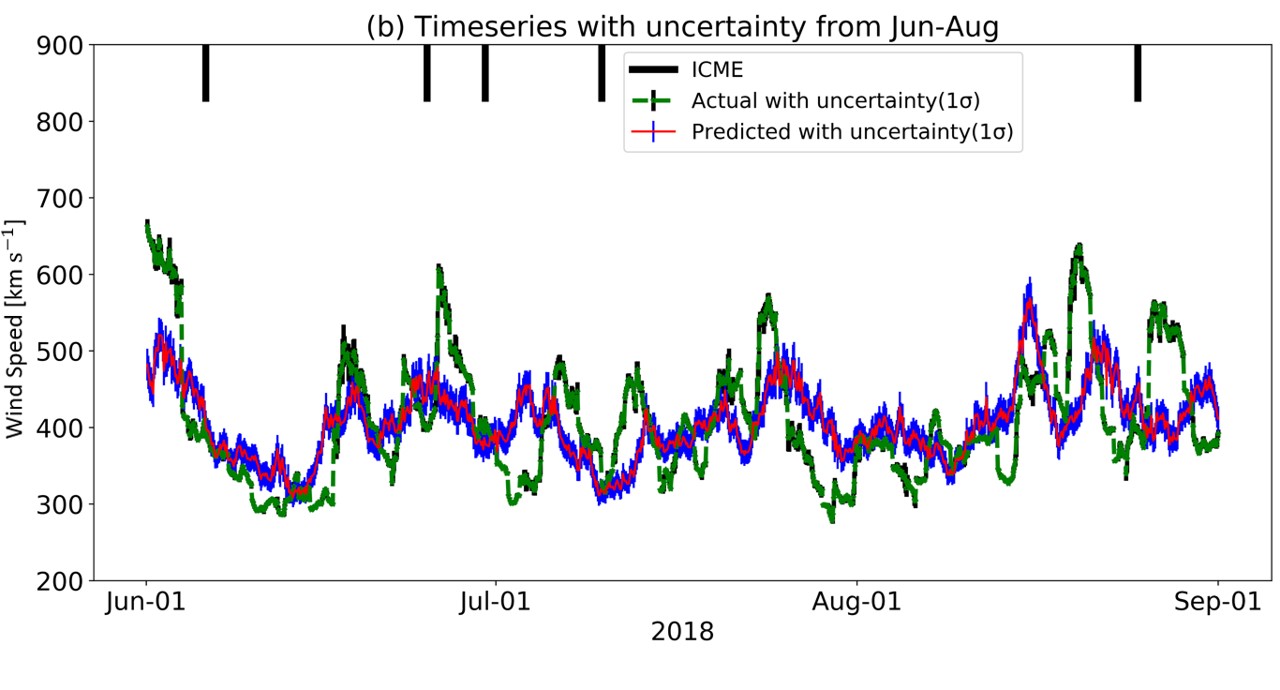}
     \caption{Time series of predicted results is shown in red with model uncertainty ($ 1 \sigma $) in blue. Observed time series is shown in green with measurement error ($ 1 \sigma $) in black. (a) Time series of predicted and actual with model and measurement uncertainty for the whole year 2018 (b) Time series of predicted and actual with model and measurement uncertainty during poor predictions.}
     \label{fig:asc}
 \end{figure}
A dropout of 0.3 before the final dense layer has been added in the present model for this purpose.\\
Dropping the nodes of the neural network randomly each time is equivalent to the model sampled from the approximate posterior distribution. Thus each sampled model corresponds to model likelihood, which helps in evaluating the predictive distribution. The use of dropout causes a perturbation to the network during each forward pass and provides a corresponding output. The same set of inputs are passed multiple times through the neural network, which generates a collective set of outputs equivalent to Monte Carlo sampling, also known as Monte Carlo dropouts. Hence evaluation of first (mean) and second (variance) moments provides us with model output and the uncertainty in the corresponding output. Here, the network is trained with the same number of epochs with fixed parameters for 100 trials for quantifying the epistemic uncertainty.\\
Figure 8a depicts the overall proposed CNN's solar-wind speed predictions for the year 2018 with uncertainty $1\sigma $ confidence intervals. A subplot for the period of June\,--\,August when the worst predictions were observed is shown in Figure 8b. It can be seen that the proposed CNN model was able to predict the trend of both fast and slow wind reasonably well. From the speed distribution of each time series, the mean of each prediction ranges from 383\,\si{\km\per\s} to 453\,\si{\km\per\s}. The standard deviation ($ 1 \sigma $) ranges from 50\,\si{\km\per\s} to 64\,\si{\km\per\s} from the mean. The average RMSE error of the model, for 100 passes is around 76.30\,$\pm$\,1.87\,\si{\km\per\s}. It is observed that within the model uncertainty, the prediction, and the actual value match.\\
\subsubsection{Benchmark Models}
We evaluated the performance of our model against few benchmark models. The metrics are evaluated on respective data of both observed and predicted time series of the year 2018 with same temporal resolution and the results are tabulated in Table 4. In this work, the below-mentioned benchmark models are taken for comparison,
\begin{itemize}
    \item
    \textbf{One-day persistence model}:\\
    Persistence model is employed with a simple forward shift in time. The one-day persistence model forecasts solar-wind speed at time $t$, as solar-wind speed at $t-1$.\\
    \item
    \textbf{27-day persistence model}:\\
    The 27-day persistence model forecasts solar-wind speed at time $t$, as solar-wind speed at time $t-27$.\\
    \item
    \textbf{Constant four-day time delay}:\\
    The four-day time delay method refers to the time delay taken as constant four days in between SDO images being captured and the wind speed arriving at the L\textsubscript{1} point. To compare the model's performance, we trained and tested the same CNN model with a constant-delay approach.\\
    \item
    \textbf{CH area regression model}:\\
    This method refers to the regression of the solar-wind speed from fractional CH area obtained from SDO images \citep{Rotter2015}. Since \cite{Bu2019} had forecast findings for the year 2018 using the \cite{Rotter2015} model, results were requested from B. Luo, to ensure fair comparison with the model.\\
\end{itemize}
\begin{table}[htp]
    \caption{Comparison of the proposed CNN model results with benchmark models}
\label{tab1}
\begin{tabular}{lp{1.5cm}p{1.5cm}p{2cm}}   
 \hline                   
\textbf{Model} & \textbf{Analysed period} & \textbf{CC} &  \textbf{RMSE [\si{\km\per\s}]} \\  
 \hline
Proposed CNN & 2018 & 0.57$\pm$\ 0.02 & 76.3$\pm$\ 1.87  \\

1d-persistence & 2018 & 0.62 & 77 \\

PS-27 \citep{Owens2013}& 2018 & 0.45 & 94 \\

CH model \citep{Rotter2015}& 2018 & 0.38 & 102  \\

Constant four-day delay  & 2018 & 0.55$\pm$\ 0.02 & 78$\pm$\ 3.6 \\

 \hline
\end{tabular}
\end{table}
The proposed CNN model gives a good overall correlation coefficient of 0.57\, $\pm$\,0.02 between predicted and actual results, when compared to standard benchmark models such as PS-27 \citep{Owens2013} for the entire year 2018. The PS-27 model correlation coefficient is around 0.45 and has RMSE of 94\,\si{\km\per\s}.\\
The histogram of the speed difference calculated between the proposed CNN, benchmark model's predictions and the actual wind speed is shown in Figure 9. From the error distribution of the prediction results, we can see that the proposed CNN model, PS-1d and PS-27 has mean around zero, whereas four-day delay and CH contour models have a shifted mean. A mean shifted to the positive side indicates that the predicted model under-forecasts the speed. A mean shifted to negative indicates that the model over-forecasts the speed. For around 290 out of 361 days, the error speed range is well within $\pm100$\,\si{\km\per\s}. This error range is also found to be in agreement with work presented by \cite{Shugay2017}. The overall percentage error for our predicted speed lies within the range of $\pm30$\,\%. 
\begin{figure}[h!]
\centering
  \centerline{\includegraphics[width=\columnwidth,clip=]{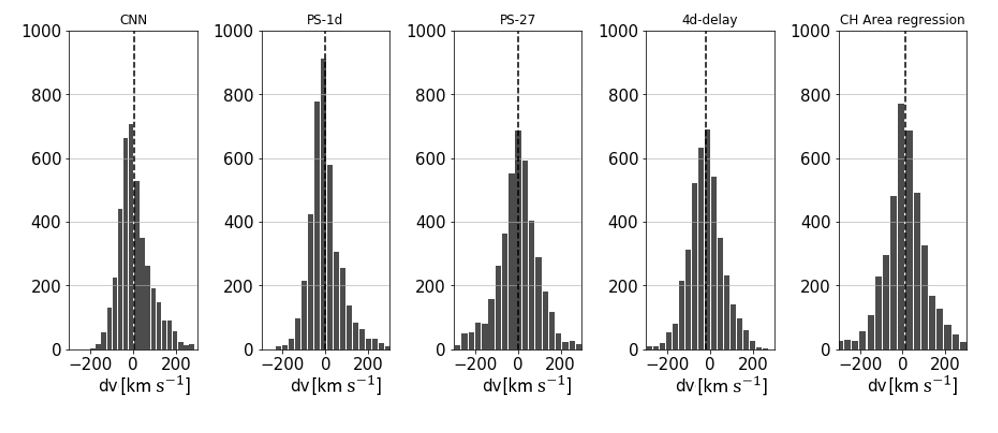}     }
              \caption{Histogram of difference in speed, calculated for CNN and benchmark models, are shown. The dotted vertical-line represents the mean of the error distribution. }
  \label{fig:asc}
\end{figure}
\begin{figure}[h!]
\centering
  \centerline{\includegraphics[width=\columnwidth,clip=]{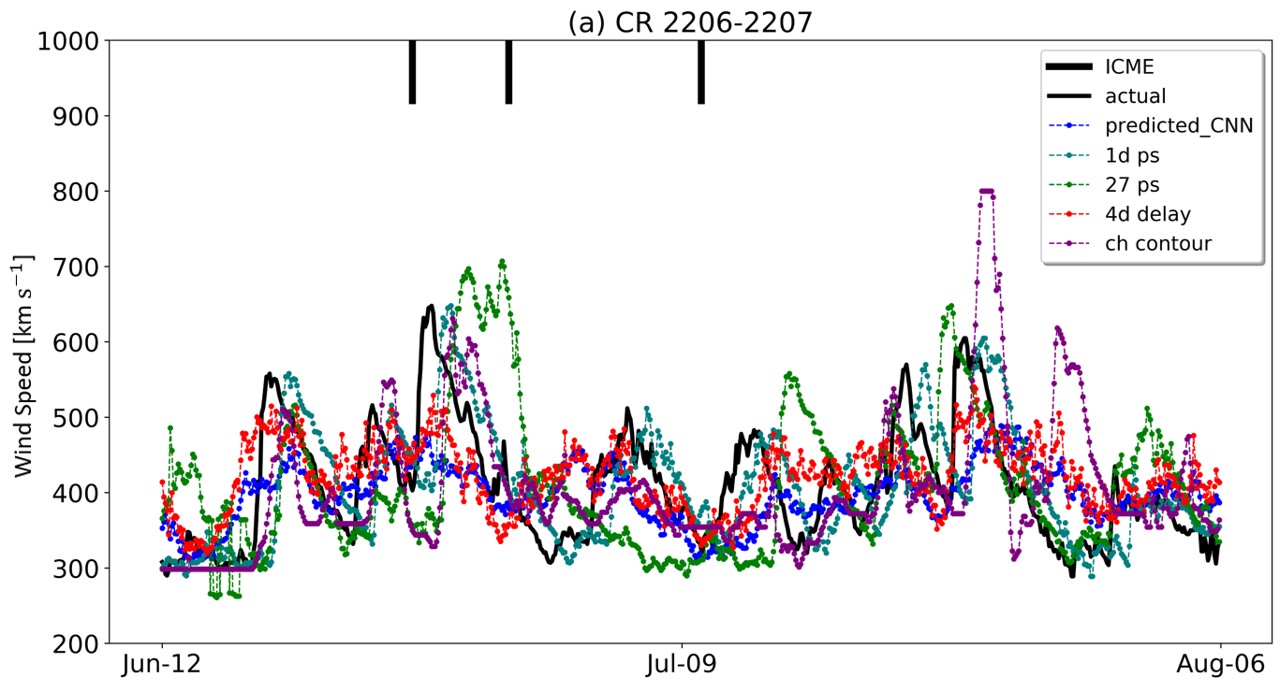}}
  \centerline{\includegraphics[width=\columnwidth,clip=]{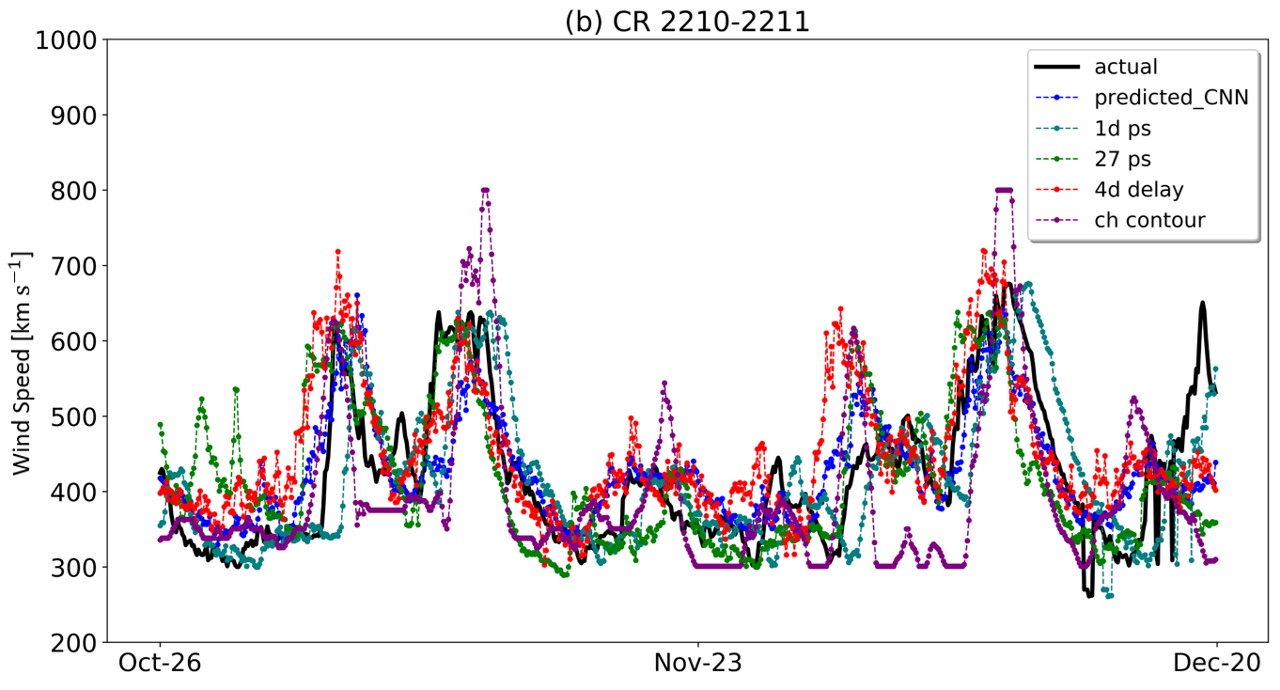}}
              
              \caption{Comparison of prediction results of CNN model with benchmark models for worst and best two carrington rotations for the year 2018. (a) shows worst prediction by CNN due to ICME disturbances during period of June to August. (b) shows best predictions by CNN during period of October to December.}
  \label{fig:asc}
\end{figure}
\subsubsection{Analysis of Results Based on Carrington Rotations}
For better understanding, evaluated performance of the proposed CNN model with benchmark models on the worst and best two carrington rotations for the year 2018 is shown in Figure 10a and 10b. From the analysis, the results for the carrington rotations 2200\,--\,2203 and 2208\,--\,2211 with little or no ICME disturbances gave a good prediction of fast and slow wind. For carrington rotations 2206\,--\,2207 poor performance by the proposed model is noticed due to frequent ICME disturbances as indicated in Figure 10a with black bars. It can be also noticed from Table 5 that correlation coefficient for every two carrington rotation overall varies between 0.21 to 0.73. For individual carrington rotations that have fewer ICME disturbances, the correlation is higher for other conventional prediction methods.\\
\begin{table}[h!]
    \caption{Correlation coefficient and RMSE of CNN model and benchmark models for every two CRs for the year 2018} 
\label{tab1}
\begin{tabular}{p{0.8cm}p{0.9cm}ccccp{0.9cm}p{0.9cm}}     
  \hline                   
Model & Para- meter & CR[1,2] & CR[3-4] & CR[5-6] & CR[7-8] & CR[9-10] & CR[11-12] \\
  \hline
CNN & CC &  0.67$\pm$0.04 &	0.66$\pm$0.04 & 0.61$\pm$0.05 & 0.21$\pm$0.07 & 0.57 $\pm$0.05 &	0.69 $\pm$0.04\\
 &RMSE &59$\pm$4.18&80$\pm$2.72&77$\pm$3.54&86$\pm$2.11&78 $\pm$2.11&71 $\pm$2.28\\
 
 PS-1d &	CC&	0.64&	0.67&	0.65&	0.60	&0.54	&0.72\\
	&RMSE&	63&	82&	79&	72	&83&	71\\

PS-27&	CC&	0.38&	0.41&	0.41&	0.41	&0.45&	0.63\\
	&RMSE&	85&	101&	115&	86&	89&82\\
 
4-day &	CC & 0.48$\pm$0.06 &	0.67$\pm$0.04 &	0.66$\pm$0.04 & 0.37$\pm$0.07 & 0.53 $\pm$0.06 &   0.63 $\pm$0.04\\
	&RMSE&	66$\pm$6.2&80$\pm$3.5&	73$\pm$4.1&	86$\pm$3.9&	75$\pm$3.1&	80$\pm$5.7\\
   
CH Area&  CC &0.08&	0.42&	0.50&	0.29	&0.61&	0.60\\
	&RMSE&	121	& 102& 89& 105& 91& 96\\
  \hline
\end{tabular}
\end{table}
\subsection{Event-Based Approach}
The metrics mentioned with continuous variables can be seen to validate a model’s basic performance. Point-to-point comparison of the variables can give us basic insight of our model’s working. However, these metrics will not indicate the performance of the model for prediction of high speed enhancements (HSEs). HSEs are identified based on sudden speed transition from slow to fast solar-wind within a short time interval. These HSEs are regarded as significant events to be detected by the forecast models.\\
In this article, we have implemented HSE event based validation technique following the work of \cite{Owens2013,Jian2015,MacNeice2009}.
We detect HSEs based on the following conditions:
\begin{itemize}
    \item Identify all points where speed difference are greater than 50\,\si{\km\per\s} from the previous day.
    \item Group the contiguous blocks as HSEs. Calculate start and end time of HSEs and discard all isolated points.
    \item If duration is less than 0.5 day then that HSE is discarded.
     \item From the point identified as HSEs, find the minimum speed within two days prior to the start time of HSE and mark it as $V_{\mathrm{min}}$.
     \item From the start time of a HSE to one day after the HSE, find the maximum speed and mark it as $V_{\mathrm{max}}$. 
     \item Find the last time reaching $V_{\mathrm{min}}$ and first time reaching $V_{\mathrm{max}}$. This duration is marked as stream interaction regions (SIRs).
     \item Regroup SIRs, find $V_{\mathrm{min}}$ and $V_{\mathrm{max}}$ for each SIR again. Eliminate redundant SIRs.
     \item SIRs with $V_{\mathrm{min}}$ greater than 500\,\si{\km\per\s} and $V_{\mathrm{max}}$ less than 400\,\si{\km\per\s} or difference between $V_{\mathrm{max}}$ and $V_{\mathrm{min}}$ less than 100\,\si{\km\per\s} are discarded \citep{Jian2015}.
\end{itemize}
Calculating hits and misses based on following conditions are shown in Table 6.
\begin{itemize}
     \item When the model predicted HSEs fall within the boundaries of actual HSEs, then it is marked as hits and termed as True Positive.
    \item When the model predicted HSEs fall outside the boundary of actual HSEs, but the start time of predicted HSE occurs within two days duration of actual HSE start time \citep{Jian2015}, then the respective HSE is also marked as hit, unless and until no HSE falls in that respective boundary of actual HSE.
    \item When there is an actual HSE event and the model does not predict any HSEs then it is considered as a miss and termed as False Negative.
    \item If no HSE event had actually occurred, but the model is predicting the event as HSE, then it is considered as false alarm and termed as False Positive.
\end{itemize}  
\begin{figure}[h!] 
    \centering
      \centerline{\includegraphics[width=\columnwidth,clip=]{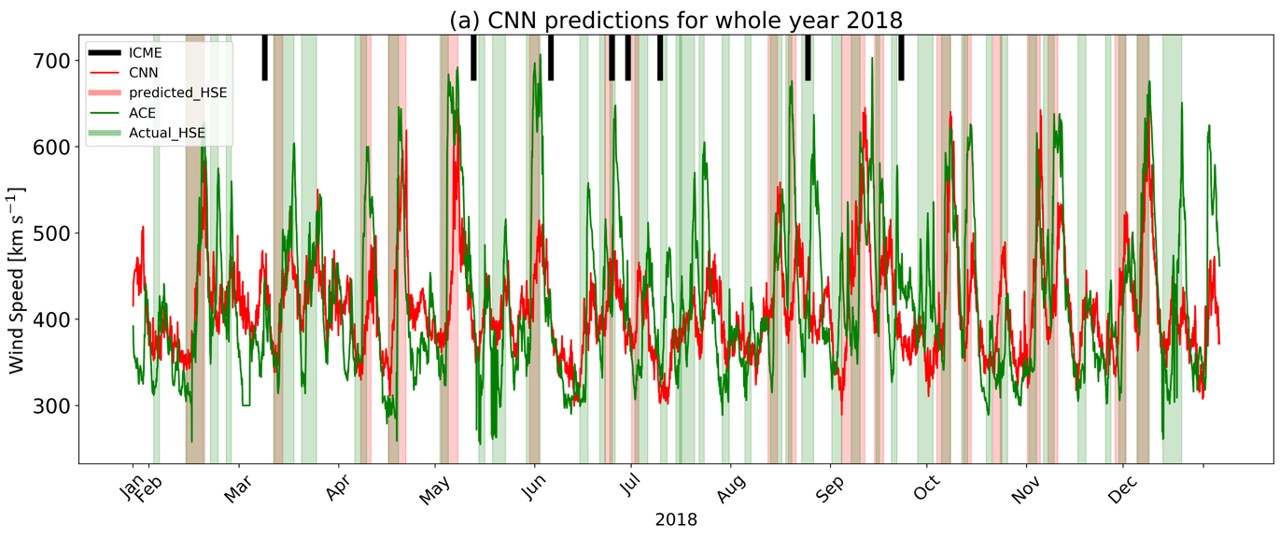}}
      \centerline{\includegraphics[width=\columnwidth,clip=]{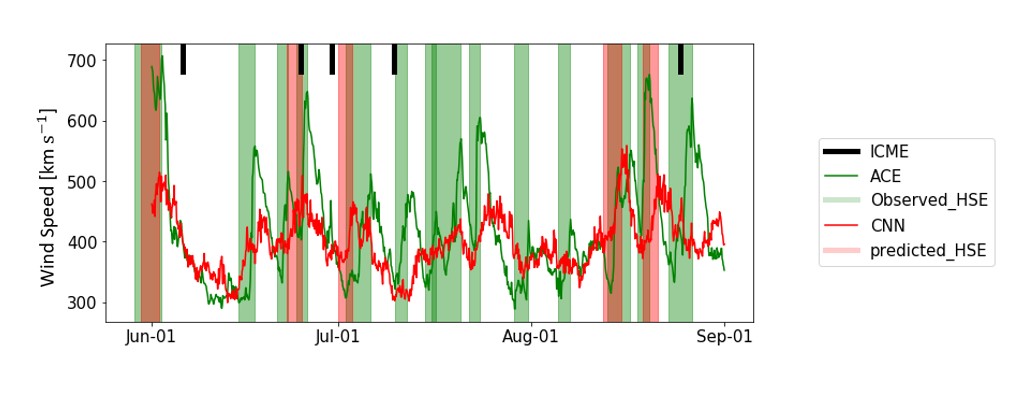}}
    \caption{(a) Predicted solar-wind speed by the proposed CNN model for the entire year 2018 are given in red, with predicted HSE events in red shading. Actual solar-wind speed taken from ACE is given in green, with its respective HSE events identified in green shading. Black bars on the top indicates occurrence of ICME events for 2018. (b) During the period of June to August, CNN had poor predictions and failed to capture HSEs due to frequent ICME disturbances.}
  \label{fig:asc}
\end{figure}
Bias of the model can be calculated as,\\
\begin{equation}
    \text{Bias} = \frac{\text{Hits+Misses}}{\text{Hits+False alarms}}\cdot
\end{equation}
The bias for CNN model comes around 2.15, which indicates that our model misses more HSE events.\\
The probability of detection (POD) will give us information on probability of the event occurrence. POD can be defined as the ratio between number of hits to number of hits and misses. POD for our model is around 46\,\% and indicates the probability that the model will identify HSEs.\\
The Positive Predictive Value (PPV) observed here, denotes that the 100\,\% of the time identified HSEs will take place. It can be calculated as the ratio of number of hits to number of hits and false alarms:
\begin{equation}
    \text{PPV} = \frac{\text{Hits}}{\text{Hits+False alarms}}\cdot
\end{equation}
False Alarm Rate (FAR) indicates the probability that the model identifies the event as an HSE when there is no actual occurence of an HSE event. FAR is given as, 
\begin{equation}
    \text{FAR} = \frac{\text{False alarms}}{\text{Hits+False alarms}}\cdot
\end{equation}
It is to be noted that the four-day model had 11\,\% FAR while PS-27 model had FAR of 16\,\%. The CH area regression model produced eight false alarms, which amounts to FAR of 20\%. However, CNN did not produce any false alarms throughout the year 2018.\\
Threat Score (TS), measures the skill of the model ranging from zero to one, with zero refers to no skill and one refers to excellent skill. TS gives information on what fraction of observed events were correctly predicted by the model.
TS can be defined as,
\begin{equation}
    \text{Threat score(TS)} = \frac{\text{Hits}}{\text{ Hits+Misses+False alarms}}\cdot
\end{equation}
From Table 6, we can see the CNN model's performance has lowered in terms of TS while identifying HSEs. We observed that the primary cause for missing of the HSEs by CNN model is due to the speed increase criterion being not achieved in few cases, and setting a threshold of 100\,\si{\km\per\s} for difference between $V_{\mathrm{max}}$ and $V_{\mathrm{min}}$ also rejects few instances. We observed that the HSE threshold kept at 100\,\si{\km\per\s} resulted in a TS of 0.46 while at 80\,\si{\km\per\s} the TS observed was 0.53 and at 50\,\si{\km\per\s} it resulted in a TS of 0.55. Given that the maximum TS by \cite{Upendran2020} is about 0.35, this may be a general feature of DL models, though the performance in the present model is better than the prior one. The results thus indicate that the HSEs which were detected in CNN, were bound to happen, though few HSEs were missed by CNN model. The respective comparison of the events and metrics with benchmark models are listed in Table 6.\\
In Figure 11a, the identified HSEs for CNN-predicted HSEs are shown in red shading, and actual HSEs are shown in green shading. From the Figure 11a, we can observe that every prediction of the HSEs in the CNN model had took place. Figure 11b shows the predictions for the month of June\,--\,August 2018, where frequent ICMEs were observed. As a result, the presence of ICMEs during this time period could have impacted the model's ability to identify HSEs.\\

\begin{table}
    \caption{Event based validation metrics comparison with benchmark models with thresholding $V_{\mathrm{min}}$\textless500\,\si{\km\per\s} and $V_{\mathrm{max}}$\textgreater400\,\si{\km\per\s}.} 
\label{tab1}
\begin{tabular}{cccccccc}     
  \hline                   
Model name & Hits & Misses & FA & TS & Bias & POD & PPV \\
  \hline
Proposed CNN  & 19&22&0& 0.46 & 2.15 & 46\,\%& 100\,\% \\
 
 PS-27 & 35&6&7&0.72 & 0.97&85\,\%& 83\,\% \\

 4d-delay&30&11&4& 0.66  &1.20&73\,\% &88\,\% \\

 PS-1d&41&0&0&1 &1 &100\,\% &100\,\% \\
 
 CH area regression&32&9&8&0.65&1.02&78\,\%&80\,\%\\
  \hline
\end{tabular}
\end{table}
\subsection{Visualization of CNN Working}
Visualization of CNN working is necessary to ensure that the model recognises the proper features from the input data. Here, this is addressed through the Gradient-weighted class Activation Mapping (Grad-CAM) technique \citep{DBLP:journals/corr/SelvarajuDVCPB16}. A similar approach was also taken by \cite{Upendran2020}. Feature maps are derived from the last convolutional layer of the network. Heatmaps of the activated features are generated by calculating the gradients of the actual outputs and computing the weighted-feature map. Grad-CAM, in simple terms, uses a heatmap to emphasize the significant features identified by CNN.
\begin{figure}[h!]
\centering
  \centerline{\includegraphics[width=\columnwidth,clip=]{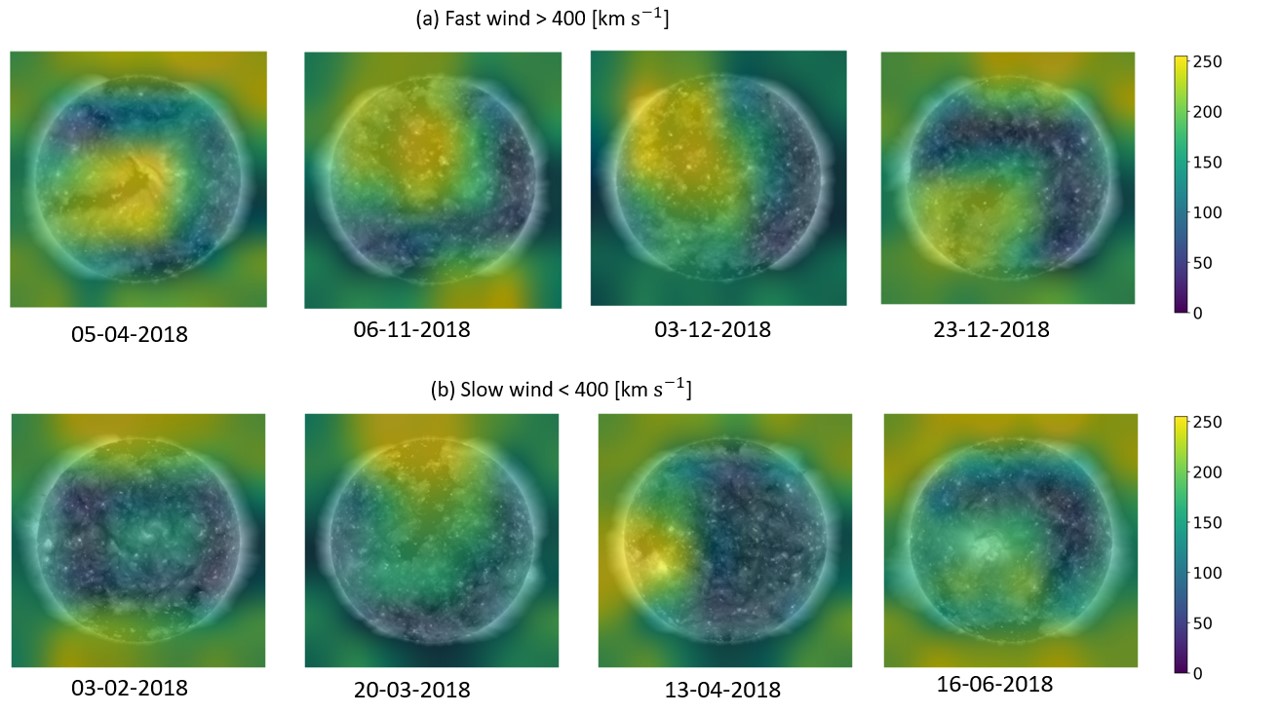}
              }
              \caption{Activation heatmap of the prediction, overlaid on the images of 2018, for fast and slow wind through Grad-CAM technique can be seen. The colorbar is mentioned in the right side of the plot, with yellow as the maximum intensity of activation. (a) shows that the maximum activation is around regions of CHs for fast wind. (b) shows that the vicinity of active regions and polar CHs are getting activated for slow wind.}
  \label{fig:asc}
\end{figure}
As an example, Figure 12 shows a few images from the year 2018, each with its heat map overlaid for both fast and slow wind. The color bar corresponds to the intensity of the activation with yellow as the maximum intensity of the heatmap. From the Figure 12a, we can observe that during fast wind, the CNN was able to activate the CH regions. It is known that CHs are the dominant source of fast wind during the descending phase \citep{Krieger1973}. Since some of the existing models are biased towards particular CH parameters, their predictions were able to generalize well mostly for the occurrence of HSEs \citep{Rotter2015}. Figure 12b shows the activation map for slow wind, where CNN displays activation near active regions and polar CHs in few cases. Several theories and observations also suggest that the slow wind might have originated from the vicinity of active regions and near coronal hole boundaries \citep{Cranmer2009}. Hence this activation analysis gives a better depiction of rationale behind CNN's performance with similar findings reported for the \cite{Upendran2020} model, for both fast and slow wind.\\
\begin{figure}[h!]
\centering
  \centerline{\includegraphics[width=6cm,clip=]{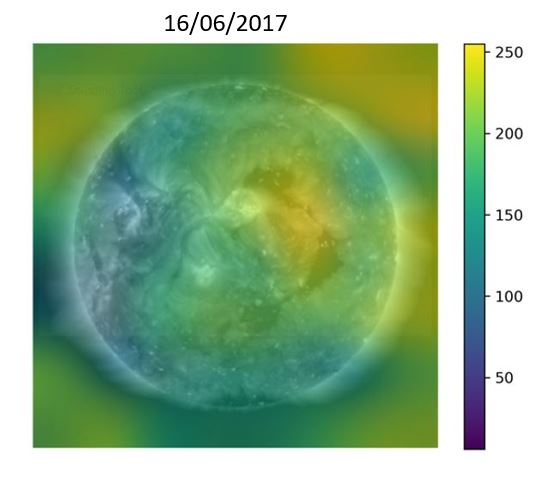}
              }
              \caption{Activation heatmap for June 16th, 2017 image, where fast and slow wind both corresponds to a single image during ballistic back tracing. It can be seen that both the CH and AR boundary regions are activated.}
  \label{fig:asc}
\end{figure}
A particular scenario of the ballistic back tracing of both fast (514\,\si{\km\per\s}) and slow wind (382\,\si{\km\per\s},355\,\si{\km\per\s}) on June 19,20 and 21 from the year 2017, resulting in mapping with the same image observed on 16th June, 2017 as shown in Figure 3. Observing the activation map by Grad-Cam visualization of that image in Figure 13, reveals that both the CH and AR boundary has been activated. However, when we tried to predict the wind speed for that image using the trained CNN model, it resulted in a speed of around 433\,\si{\km\per\s} with model uncertainty of 34\,\si{\km\per\s}. As a result, we believe that back tracing multiple speeds to a single image, might be one of the reasons that the model was not able to attain the magnitude of fast wind.\\

\section{Discussion}
\subsection{Challenges and Scope for Future Work}
The proposed model was found to give better prediction than benchmark models. We have used only one channel of AIA (193\,\si{\angstrom}) as our initial work to predict only one parameter, i.e, wind speed. This can be extended to include multiple channels and to forecast multiple solar parameters like density and temperature. The inﬂuence of ICMEs and solar flares had also affected our predictions to some extent for the months of June\,--\,August, which can be seen from Table 5. It can be noticed that the four-day delay model is performing slightly better than the proposed CNN model for the high ICME period. We believe that the adaptive-delay model may not be very appropriate in learning the propagation of ICMEs due to simple ballistic back tracing. It could be also due to the very limited number of ICME events during training of the model.\\
Here the performance of the model is evaluated based on random train--validation split. A whole separate year is then used for validating our model's performance. This method is preferred over the cross-validation approach to evaluate the model's performance for completely unseen data. Further, with a continuous one-year test data, we were able to evaluate the model's capability in identification of HSEs. However, the current study does not use data from the whole solar cycle and hence, some bias might be present in the current model. Also, there might be some bias due to the train--validation split, which might miss some of the features in one of the sets. This can be mitigated in future by including more data covering all phases of the solar cycle as and when data will be available.\\
While pretrained CNN with LSTM has already been used in earlier study for solar-wind prediction by \cite{Upendran2020}, the present model did not utilize the time-series information for prediction. This is also advantageous over the existing model, such as Persistence models \citep{Owens2013} for the same reason. The current work showed an improvement in prediction with the introduction of adaptive time-delay in the training phase. This eliminates the need for time-dependent architecture, thus making the model simple, less complex, and light-weight. However, this proposed heuristic assumption can also be considered as a drawback of this model, as the solar-wind flow path is radial with varying speed. At L\textsubscript{1} point, since the observed solar-wind will be mixed and evolved, identifying the exact structure of the solar-wind by ignoring the complex stream interactions may result in some uncertainties in arrival time. Hence accounting of these discrepancies must be studied further to improve the model. Although adaptive time-delay utilized in training improves the performance of the model, the arrival time of wind speed for the respective input image cannot be determined accurately.\\
Our model uncertainty ranges around 12\,--\,34\,\si{\km\per\s} and RMSE of the model is around 76.3\,$\pm$\,1.87\,\si{\km\per\s}. This shows the potential for further improvement of the proposed model in terms of accuracy and precision in the future. This can be enhanced by significantly improving the architecture and incorporation of diverse data. While the trained CNN predicts well the trend of occurrence of fast and slow wind, under prediction of wind-speed magnitude was observed similar to \cite{Upendran2020} model. In comparison to previous baseline models, performance of TS further reinforced the fact that the suggested CNN model captures the trend but not the high peak magnitude in few instances. By addressing the under prediction of the wind-speed magnitude, the performance of the CNN model can be improved drastically, both in terms of RMSE and TS. However, this might be a characteristic of deep-learning models, and  more research is needed to fully comprehend the capabilities of these models.\\
Mapping back fast and slow wind to the same image, can be considered as an example of the adversarial samples, which have not been extensively studied yet, when deep-learning is implemented to study the SDO images. This kind of degeneracy might easily confuse the network and may cause the samples to be misclassified. Misclassification due to degeneracy can be reduced by training two separate models independently for fast and slow-wind, and to integrate them later with some weightage. Inclusion of a more realistic wind-flow model may also help in avoiding the degeneracy problem.
Further, probing more on fast and slow-wind speed predictions separately with individual deep-learning networks might also reveal more information on solar-wind origin, which might be helpful in improving model's performance.\\
The GRAD-CAM activation study offers interesting results, such as enhanced AR and CH activation for slow wind, which was also reported by \cite{Upendran2020}. Detailed study should be carried out further, to determine whether such an observation is either physical or a result of employing the same Grad-CAM approach, which will help in better understanding the origins of slow wind.

\section{Conclusion}

In the past, CNN had been used mostly in the segmentation of CHs and classification of solar flares. The main motive behind this work is to analyze and improve the application of CNN in solar-wind prediction with adaptive time-delay employed during training. The advantage of using the CNN model is that there is no prerequisite of defining, preprocessing any specific features such as CH, nor designing any specific algorithm each time in accommodating new events. CNN adapts itself in learning diverse features and anomalies with appropriate data. Hence the model presented here is not biased towards any features.\\
Designing a CNN-based model for prediction of solar-wind speed resulted in desirable outcomes, except during those periods with ICME disturbances. The basic visualization of our CNN model shows that there is some overlap between the expected source regions from physics and those learnt by the network. This was similarly observed by \cite{Upendran2020} also, when capturing the sources of slow wind \citep{Cranmer2009}. But it does not guarantee that the network had learnt all appropriate features needed for this forecasting. Thus there is still a lot of scope in further improving the model's architecture. Also, by adjusting the hyperparameters and by addition of more data covering all phases of the solar cycle, we might improve the model's performance.\\
Instead of the conventional approach of taking an average three- to four-day time delay, we proposed a method of ballistically back tracing the image with corresponding solar-wind speed during training. This helped in improving the accuracy of the model. This also helped in removal of the time-dependent architecture, resulting in a computationally light-weight model. Optimum hyperparameters are chosen based on a trial and error method for good prediction.\\
In the absence of ICME disturbances, the proposed model forecasts the solar-wind speed better at 1\,AU. The CNN model was able to give better predictions for slow wind also, in comparison to benchmark models. Prediction of wind speed at L\textsubscript{1} is achieved for a best lag of four days prior to occurrence of HSEs. Our future work will include \textit{Helioseismic and Magnetic Imager} (HMI) magnetogram images and multiple wavelength images in training for further improvement in the accuracy of our predictions, especially during the high solar-activity periods and ICME events. \\

\section*{ACKNOWLEDGEMENT}

The authors thankfully acknowledge the use of data courtesy of the SDO/AIA science teams and \cite{Galvez_2019} for ML curated dataset. The authors also thankfully acknowledge NOAA/SWPC and ACE Science Center for providing the ACE data. The authors would also like to thank B. Luo for providing solar-wind forecast data on request. The authors would also like to express their gratitude to the anonymous reviewer for their insightful remarks, which greatly improved the technical quality and organization of our manuscript. This research uses Python packages numpy  \citep{harris2020array}, scikit-learn \citep{scikit-learn}, matplotlib \citep{Hunter:2007}, and tensorflow \citep{tensorflow2015-whitepaper}. For Grad-CAM visualization, we used Python package ELI5.

\vspace{-1.0cm}
\section*{}
\textbf{Disclosure of Potential Conflicts of Interest} The authors declare that they have no conflicts of interest.




    


\bibliographystyle{spr-mp-sola} 
\bibliography{solar_physics_2nd.bib} 

\end{document}